\documentclass[11pt, fleqn]{article}  
\usepackage{amsmath, amssymb,bm}
\usepackage{amsfonts}
\usepackage{graphicx}
\usepackage[section]{placeins}
\usepackage{multirow}
\usepackage{dcolumn}

\newcommand{\dt}{\delta t}

\newcommand{\tr}{\;\text{tr}}
\newcommand{\id}{\mathbb{I}}

\newcommand{\E}[1]{E\left[~ #1 ~\right]}
\newcommand{\condExpt}[1]{ E\left[~ #1 ~|~\Omega(t)\right]}
\newcommand{\SigmaEff}{\Sigma_{\text{eff}}}
\newcommand{\wInfty}{w_\infty}
\newcommand{\ev}{e}  
\newcommand{\bfv}{\mathbf{v}} 
\newcommand{\bfr}{\mathbf{r}} 
\bmdefine\bfeps{\varepsilon} 

\newcommand{\avg}[1]{\left< #1 \right>}

\newcommand{\figpath}{./figures}

\newcommand{\figQuality}[1]{
  \centering
  \includegraphics[width =0.48\textwidth]{#1/meanSpectrum_covariance}\hspace{1em}
  \includegraphics[width =0.48\textwidth]{#1/residual_stdDev_mean}\\[1ex]
  \includegraphics[width =0.48\textwidth]{#1/whiteningQuality_r_r_Correlation_L2}\hspace{1em}
  \includegraphics[width =0.48\textwidth]{#1/whiteningQuality_rSq_rSq_Correlation_L2}\\[1ex]
  \includegraphics[width =0.48\textwidth]{#1/whiteningQuality_rSq_rSqLag_Correlation_L2}\hspace{1em}
  \includegraphics[width =0.48\textwidth]{#1/whiteningQuality_stdDev}
}

\begin{document}
\begin{center}
{\LARGE\bf Inference on multivariate ARCH processes \\[1ex] with large sizes}\\[3ex]
\vspace*{8ex}
\parbox{0.35\textwidth}{\renewcommand{\baselinestretch}{1.0}\normalsize
{\bf Gilles Zumbach} \\[2ex]
RiskMetrics Group \\
Av. des Morgines 12 \\
1213 Petit-Lancy\\
Geneva, Switzerland\\[1ex]
gilles.zumbach@riskmetrics.com}

\vspace{3ex}
keywords: Multivariate GARCH processes, covariance matrix, white noise residuals 

\vspace*{3ex}
\today

\vspace{6ex}

\end{center}
\begin{abstract}
The covariance matrix is formulated in the framework of a linear multivariate ARCH process with long memory, where
the natural cross product structure of the covariance is generalized by adding two linear terms with their respective parameter. 
The residuals of the linear ARCH process are computed using historical data and the (inverse square root of the) covariance matrix.
Simple measure of qualities assessing the independence and unit magnitude of the residual distributions are proposed.
The salient properties of the computed residuals are studied for three data sets of size 54, 55 and 330.
Both new terms introduced in the covariance help in producing uncorrelated residuals, but the residual magnitudes are very different from unity.
The large sizes of the inferred residuals are due to the limited information that can be extracted from the empirical data when the number of time series is large,
and denotes a fundamental limitation to the inference that can be achieved.
\end{abstract}

\newpage
\section{Introduction}
The construction of multivariate models for financial time series is an important but difficult topic. 
Aiming at practical applications on todays portfolios, the number of time series $N$ should be large, from hundred(s) to thousand(s).
In this context, a parsimonious approach is crucial, and the estimation of the model parameters should be very efficient.
On the other hand, a realistic model should capture the heteroskedasticity and fat tails observed in the financial time series. 
These requirements impose a minimum complexity on the model, and essentially a GARCH type structure in order to capture the volatility clustering. 
A time dependent volatility is the key quantity for the heteroskedasticity, and the covariance matrix is the main object that needs to be studied in a multivariate context.

For multivariate GARCH, essentially two directions have been explored so far.
First, the extensions of the univariate I-GARCH process lead to a simple exponential weighted moving average (EWMA) scheme, with one parameter that characterizes the decay of the exponential memory. 
This parameter can be set {\it a priory} using a reasonable univariate estimation.  
This approach is used extensively in risk evaluation, following the original RiskMetrics methodology. 
In the same line, the covariance matrix can be a simple (equal weights) product of the time series in a given window, typically of one or a few years. 
Such an approach is used extensively in portfolio allocations.
The second direction that is found in the literature follows a multivariate GARCH structure, with a greater flexibility in the structure of the equations, but with a number of parameters that grows with $N$, typically as $N^2$ or $N^4$ (see e.g. a recent review in \cite{multivariateGARCH_review} and the references therein, or some of the original papers \cite{BollerslevEngleWooldridge,BEKK}).  
Even though these models could be better for capturing the structure of the empirical data, their complexity is way too large for practical applications with $N$ large. 

Progress have also be made recently with respect to univariate processes, in particular in finding processes that achieve a balance between accuracy, simplicity and robustness. 
The concepts needed to reach these goals are the long memory for the volatility and the linear structure for the  covariance \cite{Zumbach.LongMemory}.
An important stylized fact is that the lagged correlation for the volatility decays slowly (as a logarithm of the lag) for all financial time series\cite{Zumbach.RM2006_fullReport}, but certainly not exponentially fast. 
This is the long memory, observed empirically as volatility clusters at many time scales.
Processes with many time scales are needed to correctly model this long memory for the volatility.
The second concept is related to the linear versus affine structure of the variance equations.
If an affine term is included in the equations, it fixes the mean volatility, but this introduces two more parameters. 
The key observation is that for a volatility forecast, a linear structure is sufficient. 
Moreover, the long memory introduces a non trivial term structure in the volatility forecast (similar to the mean reversion of GARCH(1,1)).
This simplification of the process allows us to dispose of the mean volatility parameter, a quantity that is clearly series dependent. 
This approach has been used to build a better risk methodology in \cite{Zumbach.RM2006_fullReport}.

This paper explores the multivariate extensions of the (univariate) linear long memory GARCH process developed in \cite{Zumbach.LongMemory}.
It can be seen as a better version of the EWMA applied to multivariate time series.
The model is very parsimonious as it introduces only two new parameters, essentially related to the shrinkage of the correlation and to the regularization of the smallest eigenvalues in the covariance matrix.  
In a process setup, the covariance is used to transform the realized returns into the innovations (or residuals), which should be uncorrelated white noises by hypothesis. 
Measures of quality are introduced that quantify how good is the (inverse square root) covariance at producing white noise innovations. 

In \cite{Zumbach.empiricalCovariance}, the time dependent covariance and correlation matrices are analyzed in detail.
The extensive empirical investigations on three data sets allow to better understand the specificity of the multivariate problem, in particular the role of the largest eigenvalues and eigenvectors, as well as the small and possibly null eigenvalues. 
This last point is particularly critical for model inferences as the inverse square root of the covariance is needed.
For our purpose, the key property of the covariance is that the eigenvalues of the covariance matrix decrease exponentially fast toward zero.
The accumulation of very small eigenvalues is creating problems when computing the inverse volatility, even when the covariance is mathematicaly well behaved. 
A very similar problem appears in portfolio optimization in a mean-variance scheme, as the inverse of the covariance determines the optimal allocations. 
In both cases, a proper regularization should be introduced; this issue is investigated thoroughly in the empirical section. 

The structure of this paper is the following. 
The next section introduces the relevant theoretical material, and in particular the multivariate extensions of the linear long memory process.
The sample empirical correlations for the returns is presented in Section~\ref{sec:sampleReturnCorrelations}, 
and contrasted in Section~\ref{sec:sampleResidualCorrelations} with the sample correlation for the residuals.
Section~\ref{sec:whiteningResiduals} presents measures related to the whitening of the innovations.
The main kernels used to evaluate the covariance matrix (equal weights, exponential, long memory, long memory + regularization) are compared in Section~\ref{sec:comparingKernel}.
Finally, Section~\ref{sec:regularizationComparison} compares the cut-off that are commonly used in portfolio optimization versus the regularization introduced in this paper, before the conclusions.

\section{General framework}
\label{sec:generalFramework}
The largest quantitative correlations across assets are between the returns $\rho(r, r)$, the volatilities $\rho(r^2, r^2)$ and the lagged volatilities $\rho(L[r^2], r^2)$ (see Section~\ref{sec:comparingKernel}). 
Consequently, we need to select a time series process that can capture these effects. 
The model is built upon the (linear) univariate long memory ARCH process developed in \cite{Zumbach.LongMemory}. 
Thanks to its good volatility forecast and analytical tractability, this process has been used to develop a better risk methodology in \cite{Zumbach.RM2006_fullReport}. 
In particular, the process is very parsimonious as it contains only three parameters, with one parameter that characterizes the decay of the volatility lagged correlation, and two parameters that are simple cut-off. 
An extensive (univariate) empirical analysis in \cite{Zumbach.RM2006_fullReport} shows that the same parameter values can be used for {\it all financial time series}. 
If we accept these parameter values, we are left with a univariate process that does not contain any free parameters.

As the process is quadratic (for the return), the extension to a multivariate setting is straightforward. 
The financial universe is made of $N$ time series, indexed by $\alpha$ and $\beta$. 
The price $\mathbf{p}$, mapped price $\mathbf{x}$, return $\bfr$ are column vectors with $N$ components, while the covariance matrix $\SigmaEff$ has a size $N\times N$.
We assume that the daily data generating process (DGP), with a daily time increment $\dt$, is given by
\begin{eqnarray} 
    \mathbf{x}(t+\dt)         & = & \mathbf{x}(t) + \bfr(t+\dt)\\
    \bfr(t+\dt) & = & \SigmaEff(t)^{1/2}\;\bfeps(t+\dt).  \label{eq:baseModel_r}
\end{eqnarray}
The mapped price $x$ is $x(t) = \ln(p(t))$ for stock, stock indexes, FX and commodities. 
For interest rates, $x$ corresponds to the rate at a fixed time to maturity\footnote{More precisely, for the interest rate $R$, the mapped price is $x = \log(1 + R/R_0)$ with $R_0$ = 4\%. This mapping decouples the volatility from $R$, see \cite{Zumbach.RM2006_fullReport}. This mapping introduces a small correction on the returns for interest rates.}.
The return vector $\bfr$ is the return for the time horizon $\dt$ = 1 day.
The covariance matrix $\SigmaEff$ is the effective variance/covariance over the next time period $\dt$. 
The covariance should capture the heteroskedasticity of financial returns observed empirically, as well as the correlations across time series.
The residual $\bfeps$ (or innovation) is a white noise with a distribution $p(\varepsilon_\alpha)$. 
The usual hypothesis is that the innovations are independent and with unit variance
\begin{equation}
  \E{\varepsilon_\alpha(t) \;\varepsilon_\beta(t')} = \delta_{\alpha, \beta}\;\delta_{t, \;t'}.
\end{equation}
The residual distribution $p(\varepsilon_\alpha)$ is left unspecified for (most of) this study;
if needed, a Student distribution with 5 degrees of freedom gives a good description of the (univariate) empirical distribution (again, for all financial time series), as shown in \cite{Zumbach.RM2006_fullReport}.

In principle, a drift   $\bm{\mu} = \condExpt{\bfr(t+\dt)}$  can be included in \eqref{eq:baseModel_r}. 
As the empirical analysis shows, this term is small in magnitude, and is particularly difficult to estimate reliably. 
Moreover, its magnitude for a horizon of one day is small compared to the covariance.
Therefore, we chose to neglect it for this study and to concentrate on the covariance matrix. 

In general, we consider the class of (linear) processes where the covariance matrix $\SigmaEff$ is given by a bilinear function of the past returns.
The most straightforward extension of the univariate process is given by a ``cross product'' of the return vector 
\begin{equation}
   \SigmaEff(t) = \sum_{i = 0 }^{i_\text{max}} \lambda (i)\,\bfr(t-i\,\dt) \; \bfr'(t-i\,\dt)   \label{eq:sigmaEffCrs}
\end{equation}
with $\bfr$ a column vector and $\bfr'$ its transpose.
The weight for the past returns $\lambda(i)$ obeys the sum rule $\sum_i \lambda(i) = 1$. 
Common choices for the weights are equal weights (i.e. a rectangular window with equal weights), exponential weights (i.e. equivalent to an exponential moving average), and long memory weights. 
For the long memory process, the weights decay logarithmically slowly, with $\lambda(i) \simeq 1 - \log(i\:\dt)/\log(\tau_0)$.
The recursion equations used to define the $\lambda(i)$ and the parameter values are given in \cite{Zumbach.RM2006_fullReport}.
A process with long memory weights reproduces the long memory observed in the empirical lagged correlation of the volatility \cite{Zumbach.LongMemory}, with $\tau_0$ of the order of six years. 
For a volatility forecast based on \eqref{eq:sigmaEffCrs}, the long memory weights deliver consistently a better forecast than the other typical choices (equal weights or exponential).
For these reasons, the long memory weights are investigated in detail in this work.

In the class of linear equations between covariance and return squares, other extensions of the univariate process can be written.
An interesting first extension is similar to the shrinkage of the covariance matrix \cite{Ledoit-Wolf.2003,Ledoit-Wolf.2003.2}, but applied on the correlation.
A simple shrinkage of the correlation matrix, with the shrinkage parameter $\gamma$, is
\begin{equation}
   \rho(\gamma) = (1- \gamma) \rho + \gamma\;\id_N.
\end{equation}
where $\rho$ is the correlation matrix corresponding to $\SigmaEff$
\begin{equation}
 \rho_{\alpha, \beta} = \frac{\Sigma_{\text{eff},\alpha, \beta}}{\sqrt{\Sigma_{\text{eff}, \alpha, \alpha} \Sigma_{\text{eff}, \beta, \beta}}}.
\end{equation}
The rationale for using a shrinkage is to allow more fluctuations for the return  across assets  than what the historical correlation structure would impose.
The natural prior for the correlation is the identity, corresponding to the condition imposed on the residuals $\E{\varepsilon_\alpha \varepsilon_\beta} = \delta_{\alpha, \beta}$.
The corresponding equation for the shrinkage of the covariance matrix is
\begin{equation}  
   \SigmaEff(\gamma) = (1-\gamma) \SigmaEff  + \gamma \left. \SigmaEff \right|_\text{diag}   \label{eq:sigmaEffShr}
\end{equation}
where $\left. \SigmaEff \right|_\text{diag}$ is the diagonal part of $\SigmaEff$.
Essentially, this equation shrinks only the off-diagonal part by $1-\gamma$, whereas the diagonal part is given by the volatility of the respective time series. 

A second interesting extension consists in shrinking the spectrum of the covariance toward a multiple of the identity matrix
\begin{equation}  
   \SigmaEff(\gamma, \xi) = (1-\xi) \SigmaEff(\gamma)  + \xi  \avg{\sigma^2}\,\id_N.    \label{eq:sigmaEffReg}
\end{equation}
and with the mean variance across all assets defined by
\begin{equation}
  \avg{\sigma^2} = \frac{1}{N} \tr\left(\SigmaEff\right).
\end{equation}
The covariance has been defined so as to preserve the mean variance across assets
\begin{equation}
  \tr\left(\SigmaEff(\gamma, \xi)\right) = \avg{\sigma^2} 
\end{equation}
for all values of $\gamma$ and $\xi$.
It is easy to check that, if $\ev_\alpha$ is an eigenvalue of $\SigmaEff(\gamma)$, the corresponding eigenvalue of $\SigmaEff(\gamma, \xi)$ is $(1-\xi)\ev_\alpha + \xi \avg{\sigma^2}$.
In particular, the addition of the identity matrix changes the spectrum of the covariance by setting the minimal eigenvalue at $\xi\avg{\sigma^2}$. 
This modification is very similar to the regularization of the (inverse) covariance given below in \eqref{eq:regularization_fullRank}, and therefore we call $\xi$ the regularization parameter as it allows to compute a well defined inverse for $\SigmaEff$. 
The intuition for this term is to have the same number of (significant)  sources of randomness as of time series. 
In contrast, zero eigenvalues project out the corresponding source of randomness, and very small eigenvalues act similarly in practice.
The addition of the identity matrix introduces a minimal level of fluctuations corresponding to each source of randomness, that would be missing otherwise.
As the parameters for the long memory kernel are fixed (see \cite{Zumbach.RM2006_fullReport}), the process defined with the covariance~\eqref{eq:sigmaEffReg} has only the parameters $\gamma$ and $\xi$ that need be studied empirically.
Moreover, for $N=1$ the dependency on $\gamma$ and $\xi$ disappears and the variance reduces to the univariate case.

As emphasized above, the covariances \eqref{eq:sigmaEffCrs}, \eqref{eq:sigmaEffShr} and \eqref{eq:sigmaEffReg} are linear in the return squares.
It is easy to introduce an affine mean variance term in the process, for example with
\begin{equation}
   \Sigma_\text{eff, affine} = \wInfty \Sigma_\infty + (1-\wInfty) \SigmaEff(\gamma, \xi).   \label{eq:sigmaEffAff}
\end{equation}
The mean volatility, measured on an infinite time scale, is fixed by the $N\times N$ matrix $\Sigma_\infty$, whereas $\wInfty$ is the corresponding scalar coupling constant. 
Such a process has $N(N+1)/2 + 1$ parameters corresponding to the mean term and coupling, plus the possible parameters in the linear covariance.
To make contact with the more familiar univariate GARCH(1,1) process, the algebraic structure of the linear covariance $\SigmaEff(\gamma, \xi)$ corresponds to the I-GARCH(1) process, whereas \eqref{eq:sigmaEffAff} corresponds to the GARCH(1,1) process with $\wInfty \Sigma_\infty$ the additive constant that fixes the mean volatility (often denoted $\alpha_0$ or $\omega$ in the GARCH(1,1) literature).
A detailed discussion of the linear and affine process equations is given in \cite{Zumbach.LongMemory}.
For our purpose, we are interested in the inference about $\varepsilon$ at the scale of one day, whereas the crossover time to the asymptotic regime fixed by the mean volatility is of the order of several years (a multiple of the longest time scale included in the linear part). 
Therefore, it is perfectly justified to consider only the linear part in a study focused at one day, and therefore to reduce drastically the number of parameters in the model.

If a Gaussian random walk is the zero order model for the price behavior, the model above can be viewed as the next approximation.
Essentially, the process described by \eqref{eq:baseModel_r} with the covariance \eqref{eq:sigmaEffReg} or \eqref{eq:sigmaEffAff} is the simplest process that captures the major features of multivariate financial time series, namely the basic random walk of the prices, the volatility clustering (or heteroskedasticity), the return correlations, and the fat tail character of the price changes and of the innovations (with a Student distribution for $p(\varepsilon_\alpha)$).  Moreover, because of its simple mathematical structure, this basic model can be improved in many ways to capture finer effects. It is therefore also a good stepping stone for a more detailled description of the financial time series. 

The equation \eqref{eq:baseModel_r} is formulated as a process. Using historical data, we want to validate this model.
Inference about this simple process requires to invert \eqref{eq:baseModel_r}, namely
\begin{equation}  \label{eq:residual}
  \bfeps(t+\dt) = \SigmaEff^{-1/2}(t) \;\bfr(t+\dt).
\end{equation}
The residuals $\bfeps$ can then be used to compute a log-likelihood, for example.
As the covariance matrix is symmetric, the spectral decomposition is given by
\begin{equation}
  \SigmaEff = \sum_{\alpha = 1}^{N} \ev_\alpha \;\bfv_\alpha \bfv_\alpha'
\end{equation}
where the eigenvalues $\ev_\alpha = \ev_\alpha(t)$ and eigenvectors $\bfv_\alpha = \bfv_\alpha(t)$ are time dependent, and the eigenvectors $\bfv_\alpha$  are orthogonal. For convenience, the eigenvalues are ordered by decreasing values such that $\ev_\beta < \ev_\alpha$ for $\beta  >  \alpha$.  
Provided that the eigenvalues are strictly positive, the inverse square root covariance, or inverse volatility, is
\begin{equation} \label{eq:inverseVolatility}
  \SigmaEff^{-1/2} = \sum_{\alpha = 1}^{N} \frac{1}{\sqrt{\ev_\alpha}} \;\bfv_\alpha \bfv_\alpha'.
\end{equation}
For $\SigmaEff(\gamma = 0, \xi = 0)$ and for systems of practical interest, the covariance matrix is singular and some eigenvalues are null. 
This occurs always when the number of time series $N$ is larger than the memory length $i_\text{max}$ used to compute the covariance. 
In such case, the number of strictly positive eigenvalues is given by $N_\text{pos} = \min(N, i_\text{max})$.
For many practical applications, the memory length is of the order of one to two years ($i_\text{max} = 260$ to $i_\text{max} = 520$), whereas the number of time series can be of the order of thousand(s). 
However, even for systems that are non singular according to the mathematical criterion, the eigenvalues of the covariance matrix decay exponentially toward zero, leading to very large values in the inverse volatility. 
Clearly, this will impact the computed residuals. 
Therefore, except for systems of very small dimensions, the computation of the inverse volatility at $\gamma = \xi = 0$ needs to be regularized, even when $N < i_\text{max}$.

With a singular covariance, several schemes can be used to define an inverse volatility with an appropriate cut-off. 
A first possibility is to use only the leading eigenvalues, namely to invert the covariance in the leading $k$ subspace
\begin{equation} \label{eq:regularization_projected}
  \SigmaEff^{-1/2} = \sum_{\alpha = 1}^{k} \frac{1}{\sqrt{\ev_\alpha}} \;\bfv_\alpha \bfv_\alpha'
\end{equation}
The ``cut-off parameter'' $k$ is chosen so that $\ev_k > 0$. 
We call this scheme ``projected''.
A second possibility is to complement the previous operator so that it has full rank
\begin{equation}  \label{eq:regularization_fullRank}
  \SigmaEff^{-1/2} = \sum_{\alpha = 1}^{k} \frac{1}{\sqrt{\ev_\alpha}} \;\bfv_\alpha \bfv_\alpha'
	+ \frac{1}{\sqrt{\ev_{k+1}}} \sum_{\alpha = k+1}^{N}  \;\bfv_\alpha \bfv_\alpha'
\end{equation}
and we call this scheme ``full rank''.
A multiplicative constant can be inserted in front of both definitions so as to preserve the trace.
In practice, the optimal rank $k$ should be chosen large enough (see below), and this normalization constant is essentially irrelevant.

The singularity related to the inverse of a covariance matrix appears in many places in finance. 
A common example is the computation of the most efficient portfolio in a mean-variance scheme and the related definition of an efficient frontier.
Other examples are the computations of the factor loadings in a factor model, the inference in a multivariate process as explained above, or the computation of conditional Gaussian distribution in a multivariate setting.
Depending on the application, the choice of the cut-off can differ, with the ``projected'' definition being possibly the most widely used.
As the empirical analysis below shows, the ``full rank'' scheme is better.
Yet, the regularization through a modification of the covariance by using $\SigmaEff(\gamma, \xi)$ turns out to be the most efficient method.

Finally, we want to investigate how effective is $\SigmaEff(\gamma, \xi)$ at whitening the residuals. 
To this purpose, we compare the sample correlation for the returns and for the residuals.
The empirical average of $\mathbf{y}$ is denoted $\avg{\mathbf{y}}$, computed  over the sample from 1.1.2000 to 1.1.2008.
The covariance matrix between the vectors $\mathbf{y}$ and $\mathbf{z}$ is denoted $\Sigma( \mathbf{y}, \mathbf{z})$.
The correlation matrix between the vectors $\mathbf{y}$ and $\mathbf{z}$ is denoted $\rho(\mathbf{y}, \mathbf{z})$,  and evaluated with the usual Pearson formula.
The lag one operator is  $L[\mathbf{y}]$ with value $L[\mathbf{y}](t) = \mathbf{y}(t-1)$. 
Up to lag one, all the $N\times N$ correlation matrices are:
\begin{itemize}
    \item $\rho(\bfr, \bfr)$: The contemporaneous return correlation.
    \item $\rho(\bfr^2, \bfr^2)$: The contemporaneous volatility correlation.
    \item $\rho(\bfr, \bfr^2)$: The contemporaneous return/volatility correlation.
    \item $\rho(L[\bfr], \bfr)$: The lagged return correlation.
    \item $\rho(L[\bfr^2], \bfr^2)$: The lagged volatility correlation. This correlation characterizes the heteroskedasticity, but here also across time series.
    \item $\rho(L[\bfr], \bfr^2)$: The influence of the return moves on the subsequent volatility. For stocks, this is commonly called the ``leverage effect''.
    \item $\rho(L[\bfr^2], \bfr)$: The influence of the volatility on the subsequent returns.  
\end{itemize}
The correlations matrices for the residuals $\bfeps$ are computed similarly, but also depend on the regularization used to compute $\SigmaEff^{-1/2}$.

For the returns, all these correlations are very interesting to study as they summarize the information about the market structures and its dynamics.
For the residuals, if they are effectively iid, the correlation matrices should be either the identity matrix (for $\rho(\bfr, r)$ and $\rho(\bfr^2, \bfr^2)$) or zero (for all the other correlations). 
Simple overall measures of these relationships are given by 
\begin{equation} \label{eq:whiteningQuality_noDiag}
  q^2 = \frac{1}{N(N-1)} \sum_{\alpha \neq \beta} \rho_{\alpha, \beta}^2 
\end{equation}
for $\rho(\bfr, \bfr)$ and $\rho(\bfr^2, \bfr^2)$, and
\begin{equation} \label{eq:whiteningQuality_withDiag}
  q^2 = \frac{1}{N^2} \sum_{\alpha, \beta}\rho_{\alpha, \beta}^2 
\end{equation}
for the other correlation matrices.
Essentially, these whitening qualities $q(\mathbf{y}, \mathbf{z}) = q(\rho(\mathbf{y},\mathbf{z}))$ measure the independence of the pairs of residuals.
The unit variance of the residuals $\E{\varepsilon_\alpha^2} = 1$ should still be tested, and a simple measure is given by 
\begin{equation} \label{eq:whiteningQuality_covar}
    q^2(\bfeps^2) = \frac{1}{N} \sum_{\alpha} \left(\avg{\varepsilon_\alpha^2} - 1\right)^2.
\end{equation}
Empirically, the variances $\E{\varepsilon_\alpha^2}$ have a similar behavior for all $\alpha$, and an informative quantity is the mean residual variance
\begin{equation}\label{eq:meanResidualVariance}
     \frac{1}{N} \sum_{\alpha} \avg{\varepsilon_\alpha^2}.
\end{equation}

\section{The data sets}
\label{sec:datasets}
The three data sets we are using have been presented in \cite{Zumbach.empiricalCovariance}.
Essentially, they are the {\bf ICM  data set} (International Capital Market) that covers majors
asset classes and world geographical areas with $N = 340$, the {\bf G10 data set} covers the largest economies 
(European, Japan, and USA) with $N = 55$, and the {\bf USA data set} focuses on the American economy with $N= 54$.
All time series contain daily prices from January 1, 1999 to the January 1, 2008 (9 years), corresponding to a length of 2515 days. 
The in-sample investigations are done from 1.1.2000 to 1.1.2008, and one year ($i_\text{max} = 260$ business day) is used to evaluate the effective covariance $\SigmaEff$.

\section{The sample correlations for the returns}
\label{sec:sampleReturnCorrelations}
\begin{figure}[htb]
  \centering
 \includegraphics[width =\textwidth]{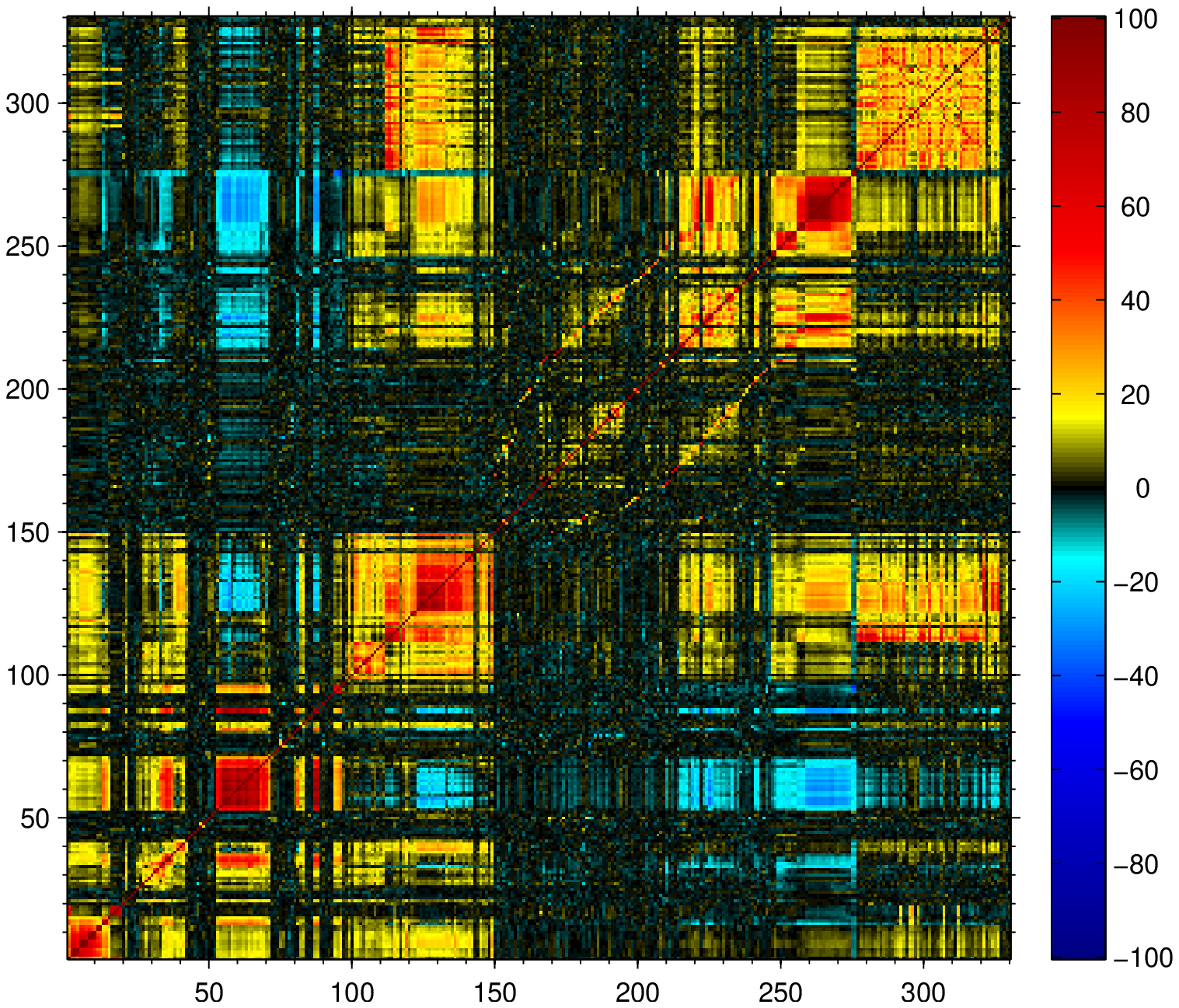}\\
  \caption{The correlation $\rho(\bfr, \bfr)$ for the returns for the ICM data set.}
  \label{fig:return_correlation}
\end{figure}
From the seven correlations between return, volatility and their lag one time series (see the end of Section~\ref{sec:generalFramework}), the largest correlation is $\rho(\bfr, \bfr)$, while the correlation between volatility $\rho(\bfr^2, \bfr^2)$ is the second largest.
The overall return correlation is presented on Figure~\ref{fig:return_correlation} for the ICM data set. 
Essentially, this figure measures the dependencies in today financial world, and several features are worth noticing on a qualitative basis.
\begin{enumerate}

\item By and large, there is one cluster of connected markets, and then disconnected areas (disconnected countries are in Asia, central America, most of South America, Central Africa, the former Soviet Union). The cluster contains all developed economies from Asia, North America, Europe and South Africa.

\item In the FX sector, there is a strongly correlated cluster around the Euro (red patch between index 53 to 70), related to another cluster containing Canada, Australia, Singapore and Japan (index 33 to 38). The rest of the Asiatic currencies are less correlated.

\item In the stock index sector, the correlations are overall fairly large, with three clusters corresponding respectively to Pacific/Asia (index 98 to 111), North America (112 to 116) and Euroland (123 to 142).

\item The interest rates show a strong difference between short maturity at 1 day (index 150 to 165) and 1 month (index 166 to 207) and long maturity at 1 year (index 208 to 245) and 10 years (index 246 to 276). 
The short maturities are essentially disconnected from all other time series, and have only small correlations with 1 year interest rates, except for a larger correlation with the same yield curve at 1 year, visible as an off diagonal line.

\item In the commodities sector, the metals (spot and future, index 1 to 15) behave very similarly to the FX, while the gas futures (index 16 to 19) have no correlations except with some American stocks in the energy and petroleum business. 

\item The correlation between European currencies and European stock indexes is negative. 
This indicates that when the dollar gets stronger (i.e. the FX rate USD/EUR decreases), the stock indexes increase.
This is probably due both to the prices of European stocks appearing cheaper to US investors, as well as American stocks appearing more expensive to European investors.

\item The IR with long maturities show clear positive correlations with stock indexes.
These correlations indicate moves in the same directions for both asset classes,  
and go somewhat against the view of alternating cycle between bonds and equities.
The time scales are however very different, as the correlations between daily price changes are computed in this paper, whereas the business cycles appear at a time scale of years and for the prices. 

\item The IR with long maturities show clear negative correlations with FX.
This is probably due to larger (smaller) US long term IR making the dollar stronger (weaker), coupled with the strong positive correlations for long terms IR.

\end{enumerate}

\FloatBarrier
\section{The sample correlations for the residuals}
\label{sec:sampleResidualCorrelations}
\begin{figure}[htb]
  \centering
  \includegraphics[width =\textwidth]{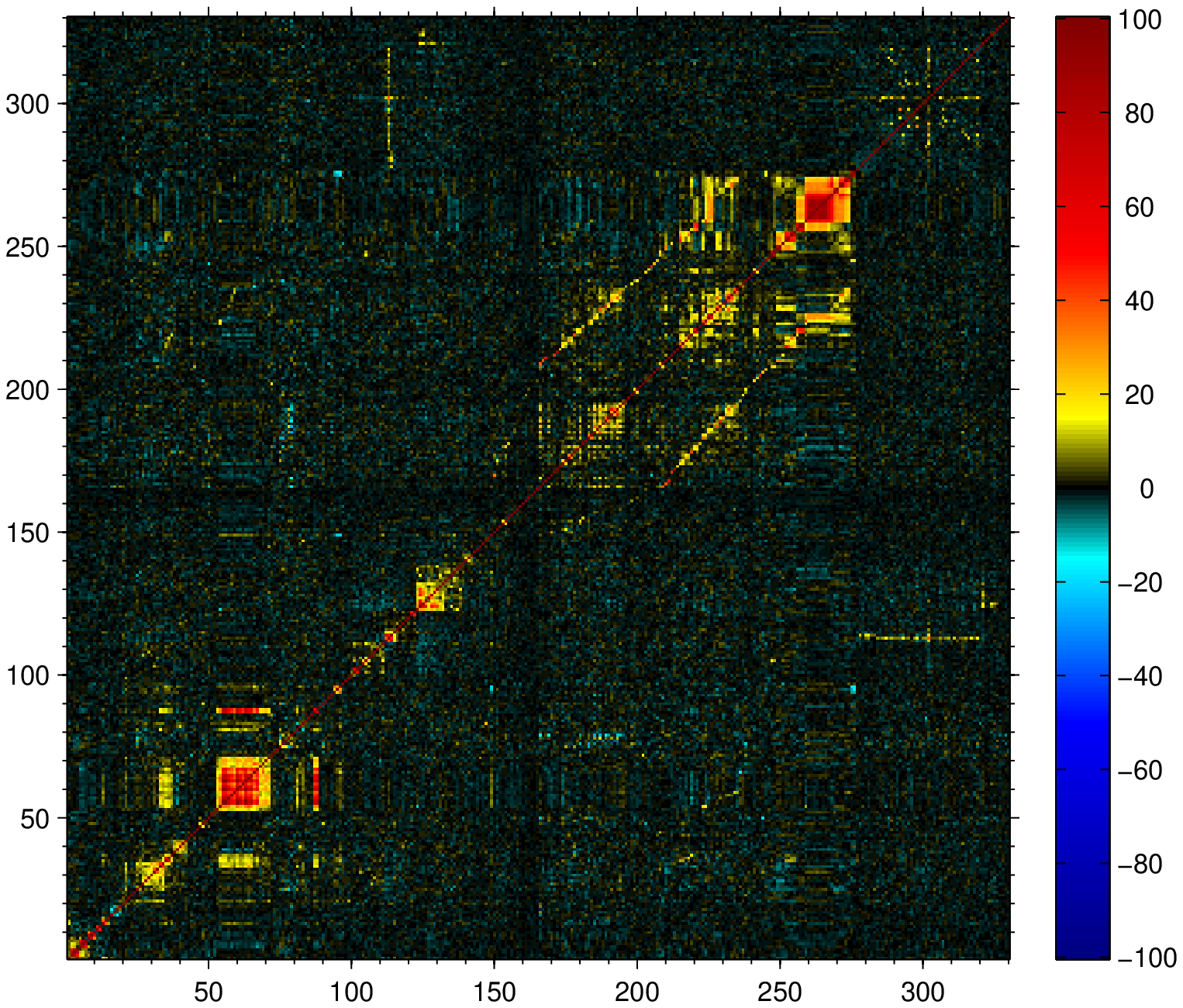}\\
  \caption{For the ICM data set, the correlation $\rho(\bfeps, \bfeps)$ for the residuals  at $\gamma = \xi = 0$ and cut-off parameter $k = 91$.}
  \label{fig:residual_correlation}
\end{figure}
The residuals $\varepsilon(t)$ are computed according to \eqref{eq:residual}, with the effective volatility $\SigmaEff(\gamma, \xi)$ given by \eqref{eq:sigmaEffReg}. 
The inverse volatility is given by \eqref{eq:inverseVolatility}.
When $\xi = 0$,  a cut-off $k$ should used in  the inverse volatility in order to avoid numerical overflows, as in \eqref{eq:regularization_projected} or \eqref{eq:regularization_fullRank}.
The empirical covariances and correlations for the residuals are computed similarly as for the returns, on the same sample (1.1.2000 to 1.1.2008).

In order to understand intuitively the statistical properties of the residuals, the correlation $\rho(\bfeps, \bfeps)$ for the residuals can be displayed, similarly to the Figure~\ref{fig:return_correlation} for the returns.
The interpretation is easier when taking  $\gamma = \xi = 0$, with a cut-off parameter $k < i_\text{max}$.
The Figure~\ref{fig:residual_correlation} corresponds to $k = 91$, with an inverse volatility computed with \eqref{eq:regularization_fullRank}.
Despite the fairly large number of eigenvalues included in the inverse volatility, structures are still clearly visible, remnant of the correlation for the returns. 
This figure makes clear that computing the inverse volatility using only the leading eigenspace for $k$ small is leaving most of the return structures in the residuals.
On the other hand, noise in the background is also visible in the form of random speckles.
The same figure plotted for increasing cut-off $k$ shows slowly disappearing structures while the background noise is increasing. 
This trade off between washing out the structures while keeping the noise small leads to an optimal choice for the cut-off $k$, or for the regularization $\xi$.
The other important criterion for the statistical properties of the residuals is that they have a unit variance.
The influence of the regularization parameter on the residual variance is large, with an decreasing variance for an increasing regularization parameter $\xi$. 
The problem is therefore to find {\it a priori} a regularization parameter that produces a unit variance for the residuals.

\section{Whitening of the residuals}
\label{sec:whiteningResiduals}
 \begin{figure}[htb]
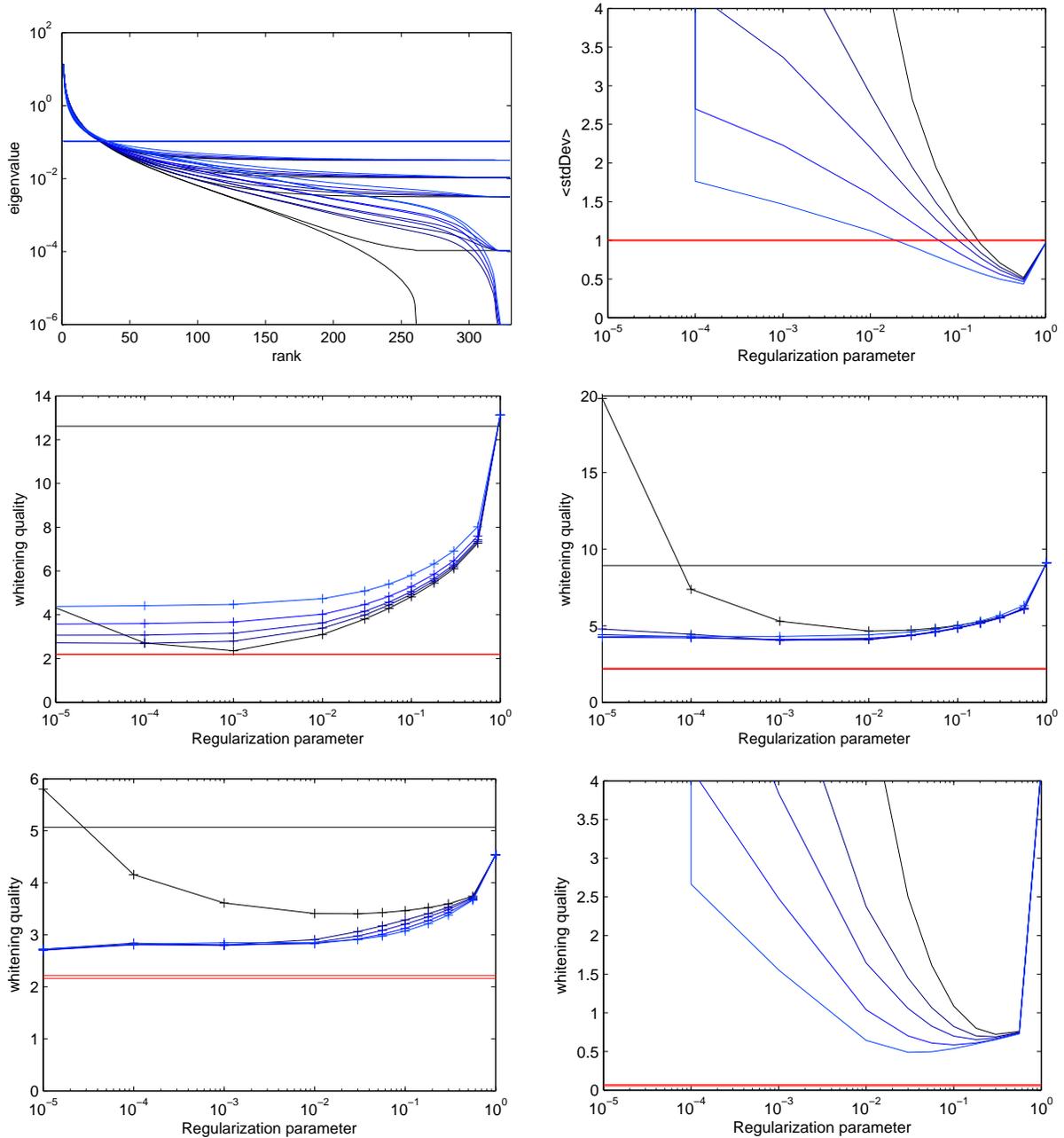

  \figQuality{\figpath/results_withRegularization/ICM}
  \caption{
	For the ICM data set, the most important whitening measures, as function of the regularization parameter $\xi$, and shrinkage parameter $\gamma$ = 0.0 (black), 0.05, 0.1, 0.2, 0.4 (blue).
	Upper left panel: the spectrum as function of the eigenvalue rank.
	Upper right panel: the mean magnitude of the residual $\avg{\bfeps}$ defined by \eqref{eq:meanResidualVariance}.
	Center left, center right and bottom left panels display respectively the whitening quality $q(\bfeps, \bfeps)$,
	$q(\bfeps^2, \bfeps^2)$ and $q(L[\bfeps^2], \bfeps^2)$.
	Bottom right: the whitening quality $q(\bfeps^2)$ for the unit variance of the residuals.
    }
  \label{fig:whiteningQuality_ICM_L2}
\end{figure}

\begin{figure}[htb]
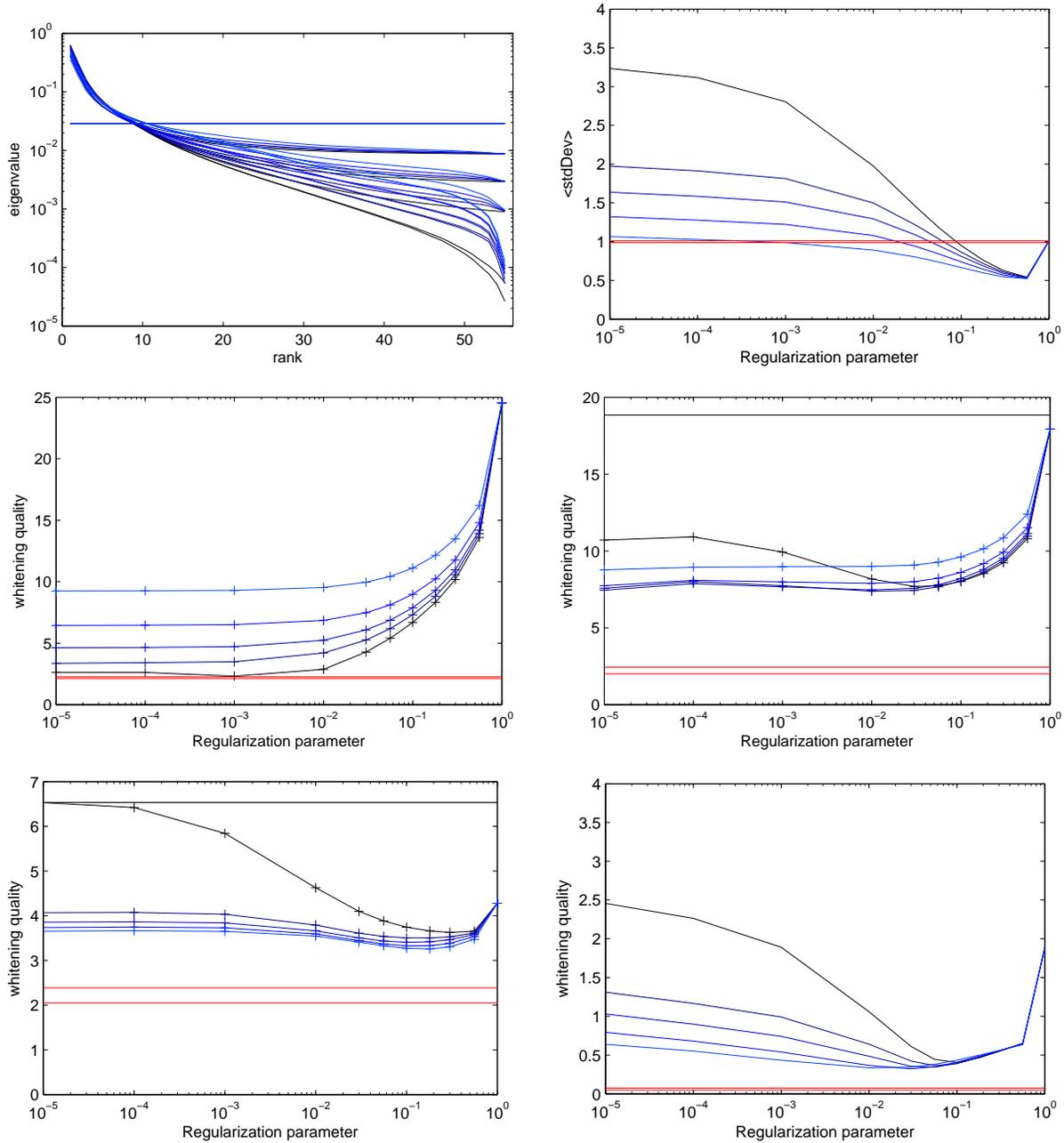

  \figQuality{\figpath/results_withRegularization/G10}
  \centering
  \caption{
	As for Figure~\ref{fig:whiteningQuality_ICM_L2}, but for the G10 data set.
    }
  \label{fig:whiteningQuality_G10_L2}
\end{figure}

\begin{figure}[htb]
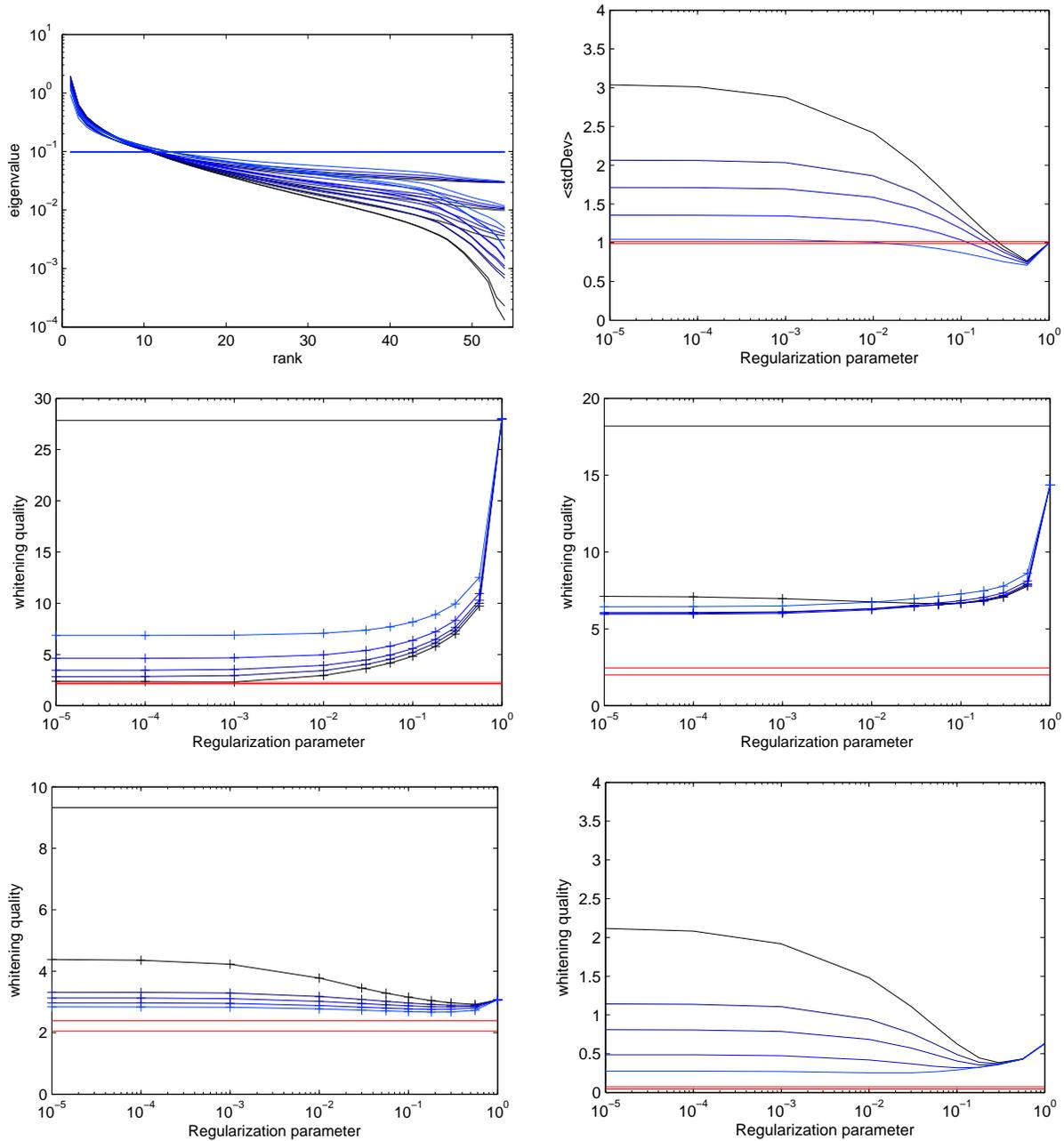

  \figQuality{\figpath/results_withRegularization/USA}
  \caption{
	As for Figure~\ref{fig:whiteningQuality_ICM_L2}, but for the USA data set.
    }
  \label{fig:whiteningQuality_USA_L2}
\end{figure}

The overall measures $q$ for the ``whitening'' of the residuals are given by \eqref{eq:whiteningQuality_noDiag}, \eqref{eq:whiteningQuality_withDiag} and \eqref{eq:whiteningQuality_covar}. 
They are plotted on Figures~\ref{fig:whiteningQuality_ICM_L2}, 
\ref{fig:whiteningQuality_G10_L2} and \ref{fig:whiteningQuality_USA_L2} for the ICM, G10 and USA data sets.
The horizontal axises give the regularization parameter $\xi$ and the black curves correspond to no shrinkage $\gamma = 0$. 
The curves with the color changing from black to blue correspond to the increasing shrinkage parameter $\gamma = 0.0, 0.05, 0.1, 0.2, 0.4$.
The  black horizontal line gives the value of the corresponding quantity for the returns, while the pair of horizontal red lines corresponds to the 5\% and 95\% quantiles of the distribution of the whitening quality for uncorrelated Student innovations (see text below). 
Essentially, the empirical whitening qualities should lies between these two extremes.
On the horizontal axis for the regularization parameter $\xi$, the point at $10^{-5}$ corresponds to $\xi = 0$ (that would be otherwise at $-\infty$ on a logarithmic axis).
For the ICM data set, the inverse volatility is computed with a floor on the spectrum of $10^{-12}$ in order to avoid numerical overflows.

For the ICM data set, the spectrum of the covariance is given on the top left panel in Figure~\ref{fig:whiteningQuality_ICM_L2}.
For this panel, the black curves correspond to $\gamma = 0$, for different values of $\xi$. 
The effect of the regularization can be clearly seen, with a spectrum that goes from singular for $\xi = 0$ (with zero eigenvalues for rank larger than 260) to a constant spectrum given by the mean volatility ($\avg{\sigma^2}\simeq 0.1 $) for $\xi = 1$.
The spectrums for increasing shrinkage $\gamma$ are drawn with colors that go from black for $\gamma = 0$ to blue for $\gamma = 0.4$.
The shrinkage effect on the small eigenvalues is very clear, with a less singular spectrum.

The mean size of the residuals are shown on the top right panel.
There is no particular feature around a unit variance, and instead the dominant feature is the very strong increase of the residual size for decreasing values of the regularization $\xi$.
Increasing the shrinkage parameter $\gamma$ alleviates the problem, but there is no plateau around $\avg{\bfeps^2} = 1$.

The whitening qualities of the residuals are displayed on the four other panels.
The largest correlations are between the contemporaneous quantities $\rho(\bfr, \bfr)$ and $\rho(\bfr^2, \bfr^2)$; the corresponding whitening qualities are plotted in the center panels.
The next largest correlation is $\rho(L[\bfr^2], \bfr^2)$, corresponding to the heteroskedasticity, and is displayed in the bottom right panel. 
For these 3 measures of quality, the best results are achieved for parameters in the range $\gamma \in [0.05, 01]$ and $\xi \in [10^{-3}, 10^{-2}]$. 
The four measures of quality related to the other correlations have a similar behavior, but with a smaller magnitude. 
The bottom right panel shows the whitening quality for the magnitude of the residuals according to \eqref{eq:whiteningQuality_covar}.
For this measure, the optimal value for the parameters are larger, with $\gamma \simeq 0.2$ and $\xi \simeq 10^{-1}$.
With this data set, it seems difficult to have optimal parameter values according to all the whitening qualities.

In order to obtain confidence boundaries around the hypothesis of no correlation for the residuals, we have used Monte Carlo simulations.
Independent residuals are drawn from a Student distribution with five degrees of freeedom, and the measure of quality computed for the same number of time series and sample length. 
This procedure is repeated 1000 times, and the 5\% and 95\% quantiles are extracted from the distribution for the measure of quality. 
Both values are plotted as horizontal red lines.
For all the measures of quality, the empirical values are clearly above the 95\%  confidence limits.
This points to the misspecification of the covariance $\SigmaEff(\gamma, \xi)$, regardless of the parameter values.
Despite this negative result, the most subtantial part of the dependencies are removed by the covariance.
For example, for the measure $q(\bfeps, \bfeps)$ (center left panel), the measure of quality for the returns is slightly below 13\%, for the residuals they are in the 3\% range, while perfectly uncorrelated residuals have a value around 2\%. 
Clearly, the less satisfactory quantitative results are for the magnitude of the residuals, with empirical values all above 0.7, while the perfectly uncorelated 95\% quantile is at 0.07. 
Let us emphasize that it is only when computing the confidence bounds that a distributional assumption is made. 

The same quantities but for the G10 and USA data sets are given respectively on Figures~\ref{fig:whiteningQuality_G10_L2} and \ref{fig:whiteningQuality_USA_L2}.
Despite the fact that these data sets relate to non singular covariance matrices and are quite different in their compositions, very similar patterns emerge compared to the ICM data set.
For both  data sets, the covariance at $\gamma = \xi = 0$ is non singular and the inverse volatility can be computed without special care. 
The whitening qualities at $\gamma = \xi = 0$ are quite diverse, with $q(\bfeps, \bfeps)$ close to an optimal value while  $q(L[\bfeps^2], \bfeps^2)$ is at the same level of $q(L[\bfr^2], \bfr^2)$, or with a magnitude for the residuals very different from 1. 
Clearly, adding shrinkage and regularization produces better overall measures of quality, with optimal values around $\gamma \in [0.05, 0.1]$ and $\xi \in [ 10^{-3}, 10^{-2} ]$.
The model with $\gamma = \xi = 0$ is clearly misspecified, while adding shrinkage and regularization improves the situation but is still not perfect.

Interestingly, the optimal values for the parameters are very similar for the three data sets, pointing to a general feature of the underlying data generating process. 
The model at $\gamma = \xi = 0$ is likely misspecified, and a process closer to the empirical data generating process is obtained with value around $\gamma \simeq 0.05$ and $\xi \simeq 10^{-3}$. 
With these values, the residuals are the closest to uncorrelated, for the three data sets.
An intuitive picture for this result can be build as follows.
With this regularization, the bottom of the covariance spectrum is at $10^{-3}\;\avg{\sigma^2}$, leading to an inverse volatility in the range $32/\sqrt{\avg{\sigma^2}}$. 
The fluctuations of the returns in the corresponding direction get multiplied by this large factor, leading to large residuals.
With an increasing matrix size, the number of very small covariance eigenvalues increases with $N$, in turn increasing the residual sizes.
In principle, the large residual problem can be solved by using large enough shrinkage and regularization parameters, but then the covariance is modified too much and leaves dependencies between residuals.

Therefore, one cannot find optimal values for the parameters that would both lead to small residual  dependencies and to unit variances. 
The later issue is essentially due the the fast decrease toward zero of the spectrum and to the possible zero eigenvalues.
Indeed, this is the signature of the limited information that can be extracted from such a multivariate system. 
Corresponding to a given kernel $\lambda(i)$, there is an effective number of time series beyond which the multivariate information gets very tenuous, as signaled by the very small or null eigenvalues. 
The obvious solution would consist in using a memory kernel that allow for more information to be harvested, but there are fundamental limitations in this direction. 
First, the long memory kernel is optimal at capturing the dynamics of the univariate volatility clustering, and using another kernel leads to worst (univariate) decorrelation of the residual. 
Then, a rectangular kernel such that $i_\text{max} > N$ is not always possible due to the limited availability of historical data. 
Moreover, such a solution would clearly miss the short term dynamics of the volatility.
Therefore, there is a fundamental limitation to the empirical multivariate information that can be extracted from a set of time series.
This limitation leads to small eigenvalues, and to the large variance for the residuals.  
Intuitively, the large residuals ``make up'' for the small eigenvalues, so that fluctuations along all possible directions do occur.
The regularization by $\xi > 0$ is such that all sources of randomness $\varepsilon$ contribute to the fluctuations of the returns. 
Notice that this is fundamentally different from a misspecification, as the process could be specified correctly, but our inference power is limited for $N$ large.

\FloatBarrier
\section{Comparing different covariance kernels}
\label{sec:comparingKernel}
\begin{table}[htb]
  \begin{center}
\begin{tabular}{| l | l | D{.}{.}{3} | llll | l |}
\hline
\rotatebox{90}{data set} & \rotatebox{90}{whitening quality}  & 
	\rotatebox{90}{return} & 
	\rotatebox{90}{Equal Weights} & 
	\rotatebox{90}{Exponential} & 
	\rotatebox{90}{Long memory} & 
	\rotatebox{90}{LM + regularization} 
	& \rotatebox{90}{white noise} \\ 
\hline
\multirow{8}{4em}{ICM}
 & $\rho(\bfeps, \bfeps)$              & 12.6 & 3.3       & 3.1          & 2.7         & 3.4 &  2.2 \\ 
 & $\rho(\bfeps^2, \bfeps^2)$     &  8.9  & 10.6     & 5.9          & 7.4         & 4.1 &  2.2 \\ 
 & $\rho(\bfeps, \bfeps^2)$          & 3.5   & 4.0       & 3.0          & 3.4         & 3.0 &  2.3 \\ 
 & $\rho(L[\bfeps], \bfeps)$         & 4.8   & 2.5        & 2.4         & 2.4         & 2.5 &  2.2 \\ 
 & $\rho(L[\bfeps^2], \bfeps)$     & 2.8   & 2.8        & 2.5         & 2.6         & 2.4 &  2.2 \\ 
 & $\rho(L[\bfeps], \bfeps^2)$     & 3.1   & 2.8        & 2.5         & 2.6         & 2.4 &  2.2 \\ 
 & $\rho(L[\bfeps^2], \bfeps^2)$ & 5.1  &  5.6        & 3.9         & 4.2         & 2.9 &  2.2 \\ 
 & $\avg{\bfeps^2} = 1$                           &         & 105       & 691        & 129        & 2.4 &  0.060 \\
\hline
\multirow{8}{4em}{G10}
 & $\rho(\bfeps, \bfeps)$              & 25.0 & 2.0  & 4.2   & 2.6  & 4.2 & 2.2 \\ 
 & $\rho(\bfeps^2, \bfeps^2)$     & 18.8 & 8.6  & 15.4 & 10.7& 7.4 & 2.2 \\ 
 & $\rho(\bfeps, \bfeps^2)$          & 5.2   & 5.0  & 6.0   & 5.2  & 4.6 & 2.7 \\ 
 & $\rho(L[\bfeps], \bfeps)$         & 6.3   & 3.6  & 3.0   & 3.2  & 3.5 & 2.2 \\ 
 & $\rho(L[\bfeps^2], \bfeps)$     & 3.2   & 3.1  & 3.4   & 3.2  & 2.7 & 2.2 \\ 
 & $\rho(L[\bfeps], \bfeps^2)$     & 3.8   & 3.3  & 3.4   & 3.3  & 2.7 & 2.2 \\ 
 & $\rho(L[\bfeps^2], \bfeps^2)$ & 6.5  & 6.1   & 7.4   & 6.5  & 3.8 & 2.2 \\ 
 & $\avg{\bfeps^2} = 1$                           &         & 0.67 &15.3 & 2.5  & 0.64 & 0.059 \\
\hline
\multirow{8}{4em}{USA}
 & $\rho(\bfeps, \bfeps)$              & 27.8 & 2.5  & 3.4   & 2.4 & 3.4 & 2.2 \\ 
 & $\rho(\bfeps^2, \bfeps^2)$     & 18.2 & 6.3  & 11.2 & 7.1 & 6.3 & 2.2 \\ 
 & $\rho(\bfeps, \bfeps^2)$          & 5.1   & 4.4  & 4.8   & 4.4 & 4.4 &  2.7\\ 
 & $\rho(L[\bfeps], \bfeps)$         & 3.6   & 2.7  & 2.5   & 2.5 & 2.5 & 2.2 \\ 
 & $\rho(L[\bfeps^2], \bfeps)$     & 3.6   & 2.7  & 2.6   & 2.6 & 2.5 & 2.2 \\ 
 & $\rho(L[\bfeps], \bfeps^2)$     & 3.9   & 2.8  & 2.6   & 2.6 & 2.4 & 2.2 \\ 
 & $\rho(L[\bfeps^2], \bfeps^2)$ & 9.3   & 4.6  & 4.5   & 4.3 & 3.2 & 2.2 \\ 
 & $\avg{\bfeps^2} = 1$                           &         & 0.48 & 10.2 & 2.1& 0.94 & 0.059 \\
\hline
  \end{tabular}
  \label{table:comparingKernel}
  \caption{The whitening qualities for the three data sets, for different kernel shapes $\lambda(i)$. 
	The column ``return'' gives the empirical values for the whitening qualities for the returns, 
        while the column ``white noise'' gives the average values for an uncorrelated Student white noise. 
	The column ``LM + regularization'' corresponds to a long memory covariance with parameters $\gamma = 0.05$ and $\xi = 0.01$. 
        For the ICM data set, the equal weights, exponential and long memory kernels are regularized using $\xi = 0.0001$.
  }
  \end{center}
\end{table}
The performances of different kernels at producing i.i.d. residuals is investigated in table~\ref{table:comparingKernel}.
All are evaluated using 260 days of history, with shapes that are equal weights, exponential with decay 0.94, long memory, and long memory with shrinkage and regularization. 
The performances of the first three kernels are somewhat similar, while the added parameters related to shrinkage and regularization makes the fourth kernel often the best. 
Yet, there is no absolute ranking.
The most salient feature is the difficulty in fulfilling the criterion $\avg{\bfeps^2} = 1$ with increasing size $N$, with the shrinkage and regularization in the fourth kernel helping significantly.
In this respect, the exponential kernel shows the worst performance, which can be understood as follow.
Due to the the fast decay of the kernel, the amount of multivariate information harvested by the exponential kernel is the smallest, leading to the fastest decays toward zero in the spectrum.
In turn, these small eigenvalues lead to the largest inverse volatilities, and therefore to the largest residual sizes. 
In contrast, the long memory kernel leads to residuals 6 times smaller, with a somewhat similar overall shape.

It would then be tempting to evaluate the correlations with a much longer kernel, for example with equal weights (corresponding to the usual formula for the correlation).
Yet, \cite{Zumbach.empiricalCovariance} finds that the correlation has a clear dynamics, with long memory.
Using a long kernel with equal weights will wash out this structure, therefore missing the short term correlation moves (this will also produces bad volatility estimators).
This  points to a fundamental limitation of the information that can be extracted from a multivariate system, when the dynamics of the correlation has also to be captured correctly.
This limitation explains the strong similarities of the key properties for the three data sets, despite that their sizes vary by a factor six, and despite that two of them are non degenerate. 
Essentially, the exponential decay toward zero of the eigenvalues (without regularization) is shared by all data sets, and this decay makes the history cut-off $i_\text{max}$ mostly irrelevant.
The exponential decay of the spectrum also explains why the shrinkage and regularization are effective at producing residuals with the desired properties. 

\FloatBarrier
\section{``Projected'' and ``full rank'' Regularization}
\label{sec:regularizationComparison}
\begin{figure}[htb]
  \centering
  \includegraphics[width =0.45\textwidth]{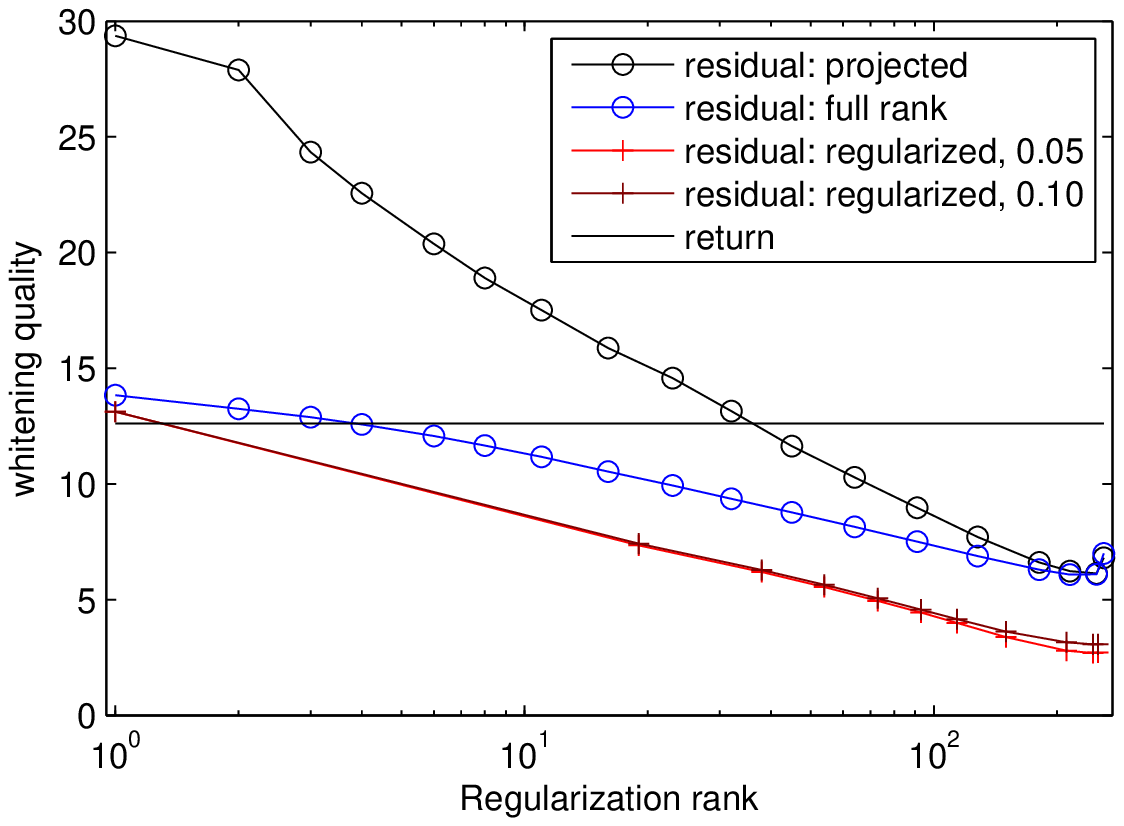}
  \hspace{2ex}
  \includegraphics[width =0.45\textwidth]{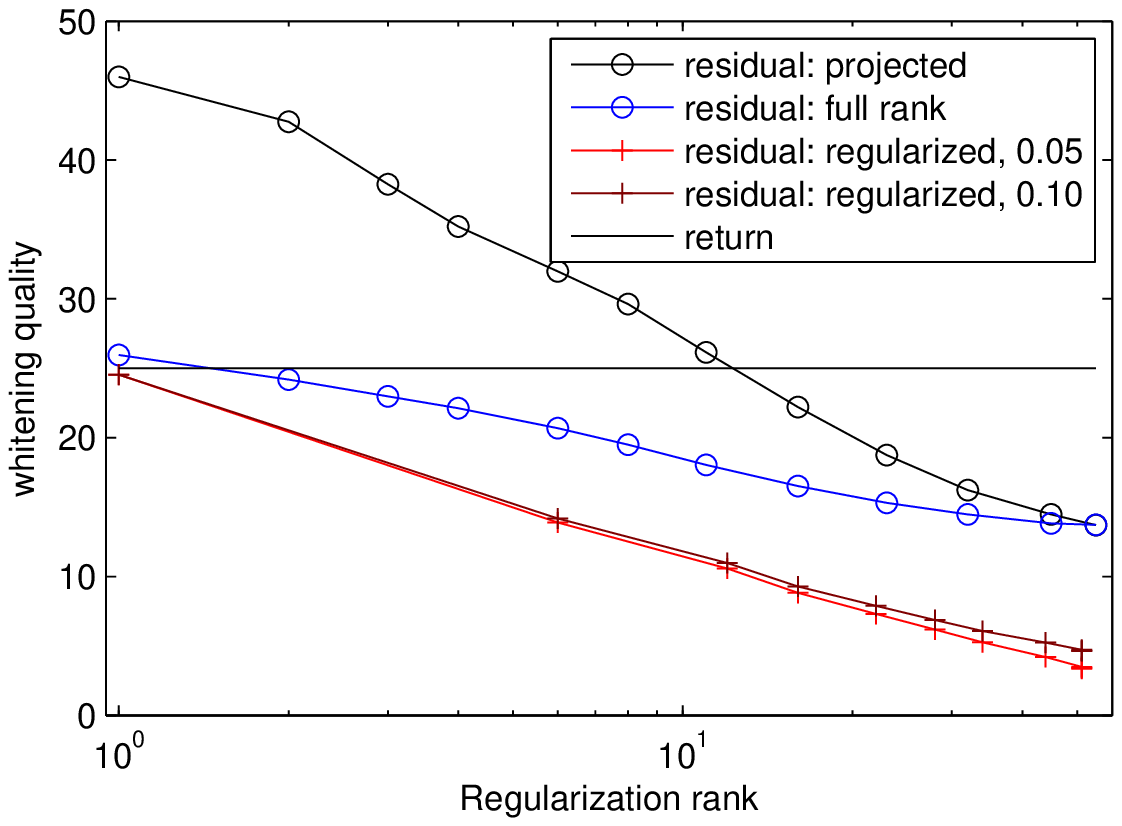}
  \caption{The whitening quality $q(\bfeps, \bfeps)$, for the ICM (left) and G10 (right) data sets.
	The regularized quality is computed with $\gamma = 0.05$ (red) and  $\gamma = 0.1$ (blue) and the regularization parameter $\xi$ is mapped to a ``plausible'' equivalent rank $k$ by using the mean spectrum. }
  \label{fig:compareProjectionVsRegularization_rho_r_r}
\end{figure}
In order to compute the inverse volatility, the most widely used method is to use the cross product covariance $\SigmaEff(\gamma = 0, \xi = 0)$, and to regularize the inverse volatility by using the ``projected'' inverse \eqref{eq:regularization_projected} or  the ``full rank'' inverse \eqref{eq:regularization_fullRank}.
It is interesting to contrast these two approaches with the regularization of the covariance with $\gamma$ and $\xi$. 
The Figure~\ref{fig:compareProjectionVsRegularization_rho_r_r} shows the whitening quality $q(\bfeps, \bfeps)$. 
The data corresponding to $\SigmaEff(\gamma, \xi)$ are mapped into an equivalent projector rank by using the mean spectrum of the covariance at $\gamma = 0, \xi = 0$. 
These curves show clearly that a ''projected'' regularization (black line) is not good for small ranks.
This finding goes clearly against a common practice consisting in using only the leading eigenvalues/eigenvectors for computing the inverse of the covariance.
The ``full rank'' regularization is clearly better than the ``projected'' scheme, for all choices of rank.
For the regularization on the covariance (red curves), the similarity of the ``full rank'' and the ``regularized'' quality is due to the similar modification of the spectrum made by both schemes.
Overall, the regularized method using  $\SigmaEff(\gamma, \xi)$ is slightly superior. 

\begin{figure}[htb]
  \centering
  \includegraphics[width =0.45\textwidth]{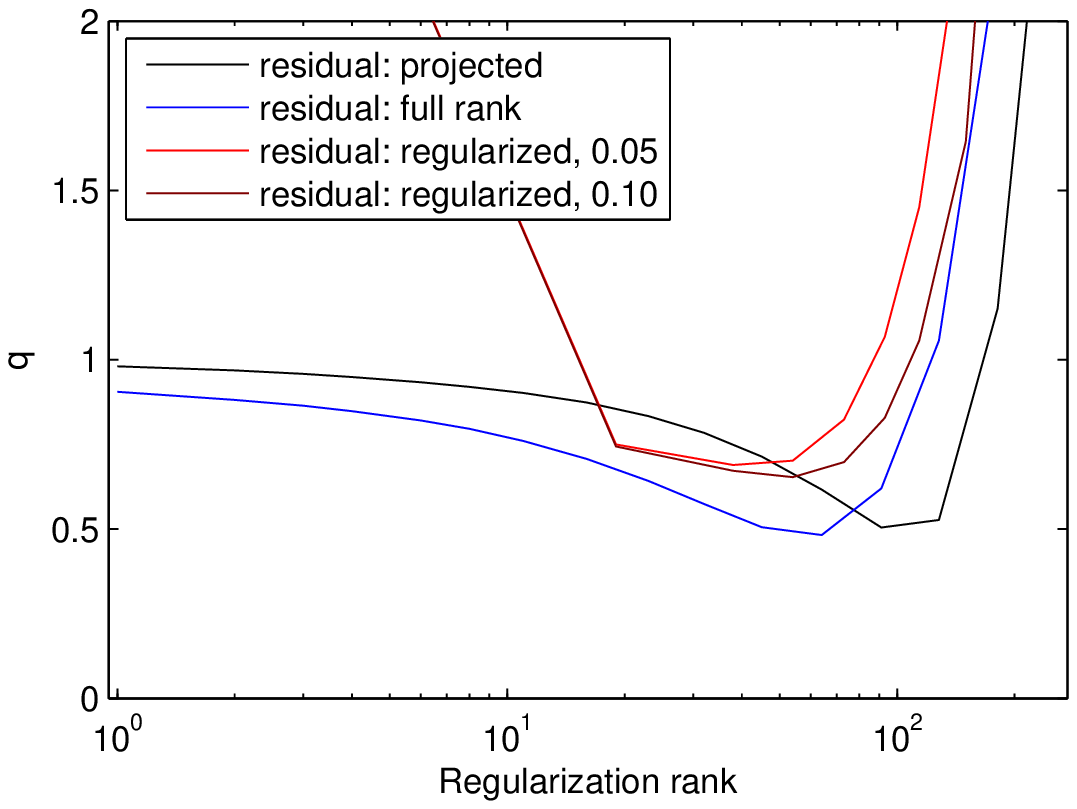}
  \hspace{2ex}
  \includegraphics[width =0.45\textwidth]{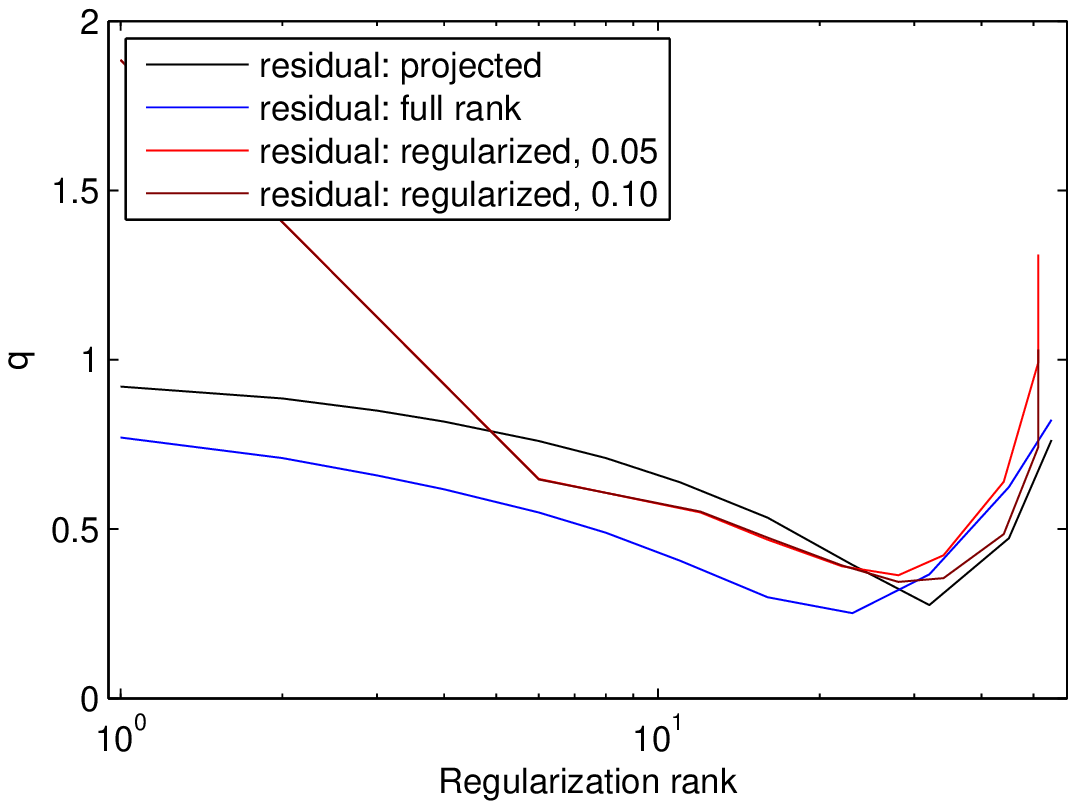}\\
  \caption{The whitening quality $q(\bfeps^2)$ for the norm of the residuals according to \eqref{eq:whiteningQuality_covar}, for the ICM (left) and G10 (right) data sets.
	The regularized qualities are computed with $\gamma = 0.05$ (red) and  $\gamma = 0.1$ (blue) and the regularization parameter $\xi$ is mapped to a  ``plausible'' equivalent rank $k$ by using the mean spectrum. }
  \label{fig:compareProjectionVsRegularization_norm}
\end{figure}
The Figure~\ref{fig:compareProjectionVsRegularization_norm} compares the measure of quality for the norm of the residuals. 
For this quality measure, the regularization of the spectrum performs better, for the ``projected'' and the ``full rank'' schemes.
However, the best measure occurs for a quite narrow range of projector ranks, and it is not clear how to select {\it a priori} the best rank.
As a first rule of thumb, a regularization parameter $k$ between 30\% to 60\% of the rank of the spectrum seems to give good results. 

\section{Conclusion}

In a multivariate process setting, the covariance defined by a simple cross product as in \eqref{eq:sigmaEffCrs} is misspecified, and a shrinking and a regularization term as in \eqref{eq:sigmaEffReg} lead to better properties for the residuals.
The spectrum of the covariance is then always non singular, and the computed residuals have statistical properties that are closer from being uncorrelated with unit variances. 

Yet, it is difficult to obtain residuals both independent and with a unit variance. 
With a given covariance kernel $\lambda(i)$, the number of very small or null eigenvalues is increasing for increasing size $N$, leading to very large values in the inverse volatility.
When multiplied by historical returns, the large inverse volatility creates large residuals, and the residual variances are increasing with $N$ (or at fixed $N$, are increasing with a decreasing cutoff as $\xi$ or $1/k$).
This difficulty is fundamentally rooted in the limited information that can be extracted from $N$ time series, leading to small eigenvalues in the covariance matrix. 
The directions corresponding to the small eigenvalues are suppressed in the process, effectively reducing the number of sources of randomness.
But in an inference computation, return fluctuations do occurs along the suppressed directions, and very large innovations are needed to compensate for the small eigenvalues.
Essentially, the limited information contained in the covariance matrix leads to a fundamental limitation on the inference that can be achieved in a large multivariate system.

This structural limitation has been diagnosed because the structure and parameters for the covariance have been inferred from a univariate linear process that has already the correct ``universal'' time structure and has no parameters (i.e. a long memory with multiple time scales instead of the more common I-GARCH(1,1) structure that has one exponential time scale).
The multivariate extension is minimal, with the two multivariate parameters $\gamma$ and $\xi$ left to be studied.

One way around the limitation would be to build a process with fewer souces of randomness than the system size $N$.
For this solution to be effective, the most relevant directions in the covariance must be stable.
For a given rank corresponding to the number of sources of randomness, the practical criterion is that the projector on the leading subspace should be stable. 
\cite{Zumbach.empiricalCovariance} shows that the projectors are instead largely fluctuating, with a long memory correlations.
Clearly, a simple approach with a fixed projector is missing part of the dynamics, and a model for the joint dynamics of the eigenvalues and eigenvectors is much more complex.

Our approach contrasts sharply with most multivariate GARCH extensions that contain of the order $N^2$ or $N^4$ parameters \cite{multivariateGARCH_review}.
The most parsimonious multivariate GARCH processes have a number of parameters of order $N$, and even this smallest number is already too large for actual portfolios.
For this reason, the multivariate extension of GARCH have been mostly confined to the academic literature. 
Beside, for a multivariate affine GARCH process, the standard procedure consists in finding the optimal parameters with a log-likelihood maximization, assuming a given distribution for the residuals. 
Following the model hypothesis, the residual distribution is set to have a unit variance in this computation.
After optimization, the empirical residuals have a variance close to one, in sharp distinction with the present results.
Indeed, the optimized parameters are ``absorbing'' the problems related to the evaluation of the large inverse volatility. 
The first effect is that the effective covariance, with the optimal log-likelihood parameters, becomes too large with respect to the volatility of the returns.
The second effect is that the parameters acquire an unwanted dependency in $N$ and that their optimal values cannot be interpreted in economic terms (say like a characteristic time for the lagged correlation). 
As a consequence, the optimal values cannot be checked for their plausibilities, and they cannot be transported from one universe to another.

As emphasized in this paper, the covariance $\SigmaEff(\gamma, \xi)$ is bilinear in the returns. 
This implies that volatility forecasts can be evaluated, namely all the quadratures can be computed analytically.
The volatility forecast at $t$ for the time $t+n\,\dt$ takes the form
\begin{equation}
   \condExpt{\Sigma_{\text{eff}; \alpha,\beta}(t+n\,\dt)} = \sum_{\alpha', \beta'} \sum_{i = 0 }^{i_\text{max}} \widetilde{\lambda}_{\alpha, \beta; \alpha', \beta'} (n, i)\,r_{\alpha'}(t-i\,\dt) \; r_{\beta'}(t-i\,\dt).   \label{eq:sigmaEffForecast}
\end{equation}
The equation for the weights $\widetilde{\lambda}_{\alpha, \beta; \alpha', \beta'} (n, i)$ is a recursion equation in $n$ that can be implemented numerically.
Therefore, the multivariate volatility forecast can be evaluated, regardless of the forecast horizon $t+n\,\dt$, for all values of $\gamma$ and $\xi$.

The present paper is partly set in a process framework, in order to justify the definition of the covariance.
The definition is however fairly general, and many shapes can be used for the weights $\lambda(i)$, including the common equal weights and exponential weights.
As found in \cite{Zumbach.empiricalCovariance}, regardless of the kernel, the spectrum decreases exponentially fast toward zero (the pace for the decay is depending on the kernel).
Therefore, a regularization of a covariance matrix computed with a given kernel --say for example a uniform kernel-- can also be achieved by adding a shrinkage and a regularization term. 
Further studies along this line, extending the works of \cite{Ledoit-Wolf.2003}, could lead to more robust results when computing an inverse variance in portfolio optimizations.

\bibliographystyle{apalike}
\bibliography{bibliography}

\end{document}